\documentclass[11pt]{article}
\usepackage{epsfig}

\newcommand{\be}{\begin{equation}}
\newcommand{\ee}{\end{equation}}
\newcommand{\ba}{\begin{eqnarray}}
\newcommand{\ea}{\end{eqnarray}}
\newcommand{\baa}{\begin{array}}
\newcommand{\eaa}{\end{array}}
\newcommand{\nn}{\nonumber \\}
\newcommand{\nr}[1]{(\ref{#1})}

\newcommand{\fr}[2]{{\frac{#1}{#2}\,}}
\newcommand{\mn}{{\mu\nu}}
\newcommand{\bfx}{{\bf x}}
\newcommand{\bfv}{{\bf v}}

\def\Tr{{\rm Tr\,}}

\def\CL{{\cal L}}
\def\CR{{\cal R}}
\def\gsim{\raise0.3ex\hbox{$>$\kern-0.75em\raise-1.1ex\hbox{$\sim$}}}
\def\lsim{\raise0.3ex\hbox{$<$\kern-0.75em\raise-1.1ex\hbox{$\sim$}}}

\begin{document}

\begin{titlepage}
\begin{flushright}
HIP-2006-59/TH\\
hep-th/0612226\\
\end{flushright}
\begin{centering}
\vfill

{\Large{\bf Spherically expanding matter in AdS/CFT}}

\vspace{0.8cm}

K. Kajantie$^{\rm a}$         \footnote{keijo.kajantie@helsinki.fi},
T. Tahkokallio$^{\rm a,b}$      \footnote{touko.tahkokallio@helsinki.fi},

\vspace{0.8cm}

{\em $^{\rm a}$%
Department of Physics,
P.O.Box 64, FIN-00014 University of Helsinki,
Finland\\}

{\em $^{\rm b}$%
Helsinki Institute of Physics,
P.O.Box 64, FIN-00014 University of Helsinki,
Finland\\}

\vspace*{0.8cm}

\end{centering}

\noindent
We discuss an exact time dependent O(3) symmetric solution with a horizon of
the 5d AdS classical gravity equations searching for a 4d boundary theory which
would correspond to expanding gauge theory matter. The boundary
energy-momentum tensor and entropy density are computed. The boundary metric
is the flat Friedmann one and
any time dependence on the boundary is incompatible with Minkowski metric.
At large times when curvature effects are negligible, perfect fluid  
behavior arises in a natural way.

\vfill
\noindent

%

\vspace*{1cm}

\noindent

\today

\vfill

\end{titlepage}

\section{Introduction}
It is of interest to search for 5d gravity duals \cite{maldacena}
of time dependent phenomena in 4d gauge theories
\cite{jp, jp2,nakamura1,nakamura2, janik}.
These could, for example, serve as prototypes of the dynamics of heavy ion
collisions. For quasi-static phenomena in strongly coupled matter
gauge-gravity duality
picture has already produced lots of interesting results, for viscosity
\cite{pss,kss,ks}, jet energy loss
\cite{lrw,hetal,buchel,gubser} and for photon production
\cite{photon}.

One can model collisions of large nuclei at very high energies by taking
the transverse size and collision energy to be effectively infinite so that
the dynamics will be invariant under longitudinal ($x_1$) Lorentz transformations.
Natural variables then are $\tau=\sqrt{t^2-x_1^2},\,\eta=1/2\log(x_1/t)$ and there
will be no dependence on $x_2,x_3$. The 5d gravity dual of this 4d physics
then has the metric \cite{jp}
\be
ds^2 =  {\CL^2 \over z^2} \left[-a(\tau,z) d\tau^2
  + \tau^2 b(\tau,z)d\eta^2+c(\tau,z)(dx_2^2+dx_3^2)+ dz^2   \right].
  \label{jpmetric}
\ee
The unknown functions $a,b$ and $c$ would be determined as solutions of
5d AdS gravity equations and 4d physics on the boundary at $z=0$ could be
computed. Several interesting
and suggestive results have been obtained by studying large-$\tau$ behavior,
without using an exact solution.
In particular, perfect fluid behavior for the energy density $\epsilon(\tau)
\sim 1/\tau^{4/3}$ is singled out by absence of curvature singularities \cite{jp},
also evidence for the viscosity/entropy density ratio $\eta/s=\hbar/4\pi$ has been
found by studying corrections to the leading large-$\tau$ behavior \cite{janik}.

To clarify the issue we will here restore the 3d spherical symmetry
broken by the metric \nr{jpmetric}. For this case we found an
exact solution, which, in fact, proved to be a known solution
\cite{langlois,kiritsis,cno} in different coordinates.
The boundary theory is the isotropic and homogeneous cosmological
FRW metric with an arbitrary scale factor $r(t)$, it is thus rather
big than little bang. In big bang energy density
decreases because space expands. Here we have no 4d gravity to determine
$r(t)$, but we shall fix it so that the comoving energy density
decreases as that in the center of a spherical little bang. Little bang
takes place in Minkowski space and the energy density in the center,
in the rest frame,
decreases because matter flows outwards. To permit one to go away from
the rest frame or the comoving frame would require a more complicated
ansatz for the 5d metric than what we use in this paper.

\section{Similarity expansion in relativistic hydrodynamics}
Experimental measurements of QCD matter in relativistic heavy ion collisions
have shown that the matter flows like a nearly ideal fluid.
As discussed above, for large nuclei
and energies the expansion can be approximated by a 1+1 dimensional
longitudinal boost invariant similarity flow of 3-dimensionally thermalised matter,
\be
v^i(t,x^1,x^2,x^3)={x^1\over t}\delta_{i1}.\quad i=1,2,3.
\label{1+1d}
\ee
Let us now generalise this to radial flow in $d-1$ spatial dimensions, keeping
$d$ arbitrary for the moment.

A radial similarity flow is defined by the velocity field
\be
\bfv(t,\bfx)={\bfx\over t},\quad u^\mu=(\gamma,\gamma\bfv)={x^\mu\over\sqrt{-x^2}},
\quad \mu=0,1,..,d-1, \quad -x^2=-\eta_\mn x^\mu x^\nu=t^2-\bfx^2\equiv\tau^2.
\label{similar}
\ee
The energy-momentum tensor is
\be
T_\mn=(\epsilon+p)u_\mu u_\nu+p\,\eta_\mn=
(\epsilon+p){x_\mu x_\nu\over \tau^2}+p\,\eta_\mn
\label{emtensor}
\ee
and
the conservation equation $\partial_\mu T^\mn=0$, split in components parallel
and perpendicular to $u^\mu$, can be written in the form
\ba
x^\mu\partial_\mu\epsilon+(d-1)(\epsilon+p)&=&0,\label{epseq1}\\
\left(\eta^\mn-{x^\mu x^\nu\over x^2}\right)\partial_\mu p&=&0.
\ea
The second is solved by $p(t,\bfx)=p(\tau)$ and, if the conformally invariant
equation of state $p=\epsilon/(d-1)$ is assumed, the first one becomes
\be
\tau\epsilon'(\tau)+d\,\epsilon(\tau)=0\quad\Rightarrow \quad\epsilon(\tau)=
{\epsilon_0\over\tau^d}={\epsilon_0\over(t^2-\bfx^2)^{d/2}}.
\label{epseq}
\ee
The constant $\epsilon_0$ here is given by the initial conditions of the flow.
Formally, it could also be fixed by including a source term on the RHS
of \nr{epseq1}.

At any fixed time $t$ the flow thus consists of a spherical shell with infinite
energy density moving radially with light velocity. In the interior the flow
pattern is of the similarity type \nr{similar} with energy density only depending
on the proper time. In the local rest frame $\bfx=0$
\be
\epsilon(\tau)=\epsilon(t)={\epsilon_0\over t^d}.
\label{restframeeps}
\ee
Our goal thus is to find the gravity dual of radially expanding thermalised
matter with the energy-momentum tensor \nr{emtensor} and comoving energy
density \nr{restframeeps}.

One may try to express $T_\mn$ in terms of radial rapidity coordinates
$(\tau,\eta_r,\theta,\phi)$, $t=\tau\cosh\eta_r,\,r=\tau\sinh\eta_r$;
$\theta,\phi$ are the usual spherical angles. Then $x^\mu\,\to\,(\tau,0,0,0)$
but $\eta_\mn\,\to\,g_\mn$ with complicated non-diagonal terms.

In the above matter is flowing in Minkowski space; compare it now with space
expanding as in the flat FRW metric ${\rm diag}(-1,r^2(t),..,r^2(t))$. Then
\be
\epsilon'(t)+{(d-1) r'(t)\over r(t)}(\epsilon+p)=0.
\label{frwflow}
\ee
One observes that if $r(t)\sim t$ this coincides with \nr{epseq1}
evaluated in the local rest frame $\bfv=0,\,\bfx=0$. For this behavior of
the radius the energy density decreases in the same way for radial flow in
Minkowski space and for Hubble flow in the FRW metric. We shall later in
Section \ref{entro} show how gauge-gravity duality selects just this
power. This is in analogy to the selection \cite{jp} of the adiabatic power
$\epsilon\sim 1/\tau^{4/3}$ for the flow \nr{1+1d}.

For completeness, the standard dissipative tensor for the flow \nr{similar}
is
\be
\Delta T_\mn=-\zeta {d-1\over \sqrt{-x^2}}\left(\eta_\mn-{x_\mu x_\nu\over x^2}\right).
\label{dissipative}
\ee
The shear part of the dissipative tensor vanishes and thus studying this flow
does not give a handle on the shear viscosity. Eq.\nr{epseq} becomes
$\tau\epsilon'(\tau)+d\,\epsilon(\tau)-\zeta(d-1)^2/\tau=0$, but
for a massless conformal fluid further $\zeta=0$.

\section{The exact solution}
In this article we shall search for
gravity duals of spherically expanding systems; the perfect fluid
solution was presented in detail in the previous section. The gravity
solution thus has to be of the type
 \ba
  ds^2 &=&   {\CL^2 \over z^2} \left[-a(t,z) dt^2
  + b(t,z)(dx_1^2 +dx_2^2+dx_3^2)+ dz^2   \right] \nn
  &=&g_{MN}dx^M dx^N\equiv{\CL^2 \over z^2}\left[g_\mn(x,z)dx^\mu dx^\nu+dz^2\right],
   \label{bulkmetric}
 \ea
where the metric is determined from the 5d AdS gravity equations
 \be
  \CR_{MN}-{1\over2}{\cal R}\,g_{MN}-{6\over\CL^2}g_{MN}=0,\quad x^M=(t,x^1,x^2,x^3,z).
  \label{ads5}
 \ee
Here the AdS radius $\CL$ is related to the 5d Newton's constant by
\be
{\CL^3\over G_5}={2N_c^2\over\pi}.
\label{magnL}
\ee

The functions $a(t,z)$ and $b(t,z)$ in \nr{bulkmetric} can now be
solved from \nr{ads5} as follows \cite{headrick}.
Using the $tz$-component of \nr{ads5}, one can solve $a(t,z)$ in terms
of an arbitrary function $F_1(t)$:
\be
a(t,z)=\frac{F_1(t)[\partial_t b(t,z)]^2}{b(t,z)}.
\ee
Using this solution for $a(t,z)$, one
can solve the function $b(z,t)$ from the $tt$-equation in terms
of two more arbitrary functions:
\be
b(t,z)=F_2(t)z^2+\frac{F_3(t)}{z^2}-\frac{1}{8F_1(t)}.
\ee
Finally, using the $zz$-equation one gets a differential
equation for $F_1(t)$, $F_2(t)$ and $F_3(t)$, from which one of the
functions can be eliminated.
We specify the remaining two arbitrary functions $r(t)$ and $h(t)$ so that
the 4d boundary metric is
\be
g^{(0)}_\mn={\rm diag}(-h^2(t),r^2(t),r^2(t),r^2(t))
\label{bound}
\ee
and the constant of integration $\sqrt2 z_0$ so that it is the position of the horizon in $z$
when $r=h=1$. One can trivially set $h=1$ by a choice of time coordinate and the
boundary metric then is a flat cosmological FRW metric.

The solution is,
abbreviating $r=r(t),h=h(t),r'=r'(t), r''=r''(t), h'=h'(t)$,
\ba
a(t,z)&=&{h^2r^2\over b(t,z)}\left\{
1+\left({h'r'\over 2h^3r}-{r''\over 2h^2r}\right)z^2+\right.\nn
&&\qquad\qquad\qquad\left.
+\left({r'^2r''\over 8h^4r^3}-{r'^4\over16h^4r^4}-{h'r'^3\over8h^5r^3}
-{1\over 4r^4z_0^4}
\right)z^4\right\}^2\nn
&\underbrace{=}_{h=1}&{r^2\over b(t,z)}\left[\left(1-{r''\over4r}z^2\right)^2-
\left({r''\over4r}-{r'^2\over4r^2}\right)^2z^4-{1\over4r^4z_0^4}z^4\right]^2,
\label{5dsola}\\
b(t,z)&=&r^2\left[\left(
1-{r'^2\over 4h^2r^2}\,z^2\right)^2+{1\over 4r^4z_0^4}z^4
\right].
\label{5dsol}
\ea

For $r(t)=h(t)=1$ the solution reduces to the metric \cite{jp}
\be
ds^2 =   {\CL^2 \over z^2} \left[-{(1-z^4/(4z_0^4))^2\over 1+z^4/(4z_0^4)} dt^2
  + \left(1+{z^4\over4z_0^4}\right)(dx_1^2 +dx_2^2+dx_3^2)+ dz^2   \right]
\label{jpmetric1}
  \ee
which, choosing a new variable $\tilde z^2= z^2/(1+ z^4/4z_0^4)$, can be brought to the
standard AdS black hole form
\be
ds^2 =   {\CL^2 \over \tilde z^2} \left[-(1-{\tilde z^4\over z_0^4}) dt^2
  + dx_1^2 +dx_2^2+dx_3^2+ {1\over 1-\tilde z^4/z_0^4}d\tilde z^2   \right].
  \label{bhmetric}
  \ee
The metrics have a horizon at $z=\sqrt2z_0, \tilde z=z_0$, with Hawking temperature
$T_H=1/(\pi z_0)$ and with entropy density $A/(4G_5V_3)=\CL^3/(4G_5z_0^3)=
\pi^2N_c^2T_H^3/2$.

\section{Energy-momentum tensor}
To study the dynamics of the boundary theory, we have to find its covariantly conserved
energy-momentum tensor. We present two methods for doing this.

For the first \cite{skenderis}, expand the $g_\mn$ in \nr{bulkmetric}, \nr{5dsola} and
\nr{5dsol} near the boundary:
\be
g_\mn(t,z)=g^{(0)}_\mn(t)+g^{(2)}_\mn(t)z^2+g^{(4)}_\mn(t)z^4+...,
\label{gexp}
\ee
where $g^{(0)}$ is the FRW metric in \nr{bound} and the rest are easy to work out from \nr{5dsola}
and \nr{5dsol}. Then the energy-momentum tensor can be evaluated using
\be
T_\mn=\frac{\CL^3}{4\pi G_5}\left[g^{(4)}_{\mu\nu}-{1\over8}g^{(0)}_\mn[(\Tr g^{(2)})^2-\Tr
   (g^{(2)})^2]-{1\over2}(g^{(2)}g_{(0)}^{-1}g^{(2)})_\mn+ {1\over4}(\Tr g^{(2)})\cdot
   g^{(2)}_\mn \right]
\ee
with the result
\be
T^\mu_{\,\,\nu}=g_{(0)}^{\mu\alpha} T_{\alpha\nu}=
{\rm diag}(-\epsilon(t),T_1^1(t),T_1^1(t),T_1^1(t)),
\label{tmunu}
\ee
where, using \nr{magnL},
\ba
\epsilon(t)=T_{tt}&=&{3N_c^2\over 8\pi^2}\left({1\over z_0^4r^4}+{r'^4\over4h^4r^4}\right),
\label{eps}\\
T^1_1(t)&=&\fr13\epsilon(t)+{N_c^2\over 8\pi^2}{r'^2(h'r'-hr'')\over h^5r^3}.
\label{epsp}
\ea
The trace of $T_\mn$ is
\be
T^\mu_\mu=g_{(0)}^\mn T_\mn=-\epsilon+3T_1^1={3N_c^2\over 8\pi^2}{r'^2(h'r'-hr'')\over h^5r^3}
\label{anom}
\ee
which is just the standard trace anomaly \cite{hs}. Further,
the tensor $T_\mn$ is covariantly conserved, $\nabla^\mu T_\mn=0$, leading for
$\nu=0$ to
\be
\epsilon'(t)+{3r'\over r}[\epsilon(t)+T_1^1(t)]=0.
\label{tcons}
\ee

For the second, the same result can be obtained without the expansion of $g_\mn(x,z)$ by
noting that, (a) due to the imposed symmetries $T_\mn$ must have the form \nr{tmunu},
(b) $T_\mn$ must be conserved as in \nr{tcons}, (c) the trace anomaly is $\CL^3/(64\pi G_5)
(R_\mn R^\mn-R^2/3)$, which leads to \nr{anom}. Eliminating $T_1^1$ using the anomaly
equation leads to an equation for $\epsilon(t)$ which can be solved to again give the result
in \nr{eps} and \nr{epsp}.

The boundary energy-momentum tensor can be naturally interpreted
to describe massless fluid in a curved background metric. The
energy density has two components:
$\epsilon(t)=\epsilon_0(t)+\Delta\epsilon(t)$. The
temperature dependent first part
$\epsilon_0(t)\sim r(t)^{-4}$ is the standard behavior of
homogeneous radiation in expanding spacetime and
the temperature independent part
$\Delta\epsilon(t)\sim r'(t)^4/r(t)^4$ describes quantum
corrections to the matter due to the curved background.

Also the $T_{ij}$ can be suggestively composed to two parts:
$T_{ij}=p(t)\delta_{ij}-\Delta T_{ij}$, where
$p(t)=1/3\,\epsilon(t)$ is the pressure of the fluid. One may try
to interpret $\Delta T_{ij}$ in terms of bulk viscosity: comparing
\nr{epsp} with the dissipative part
(Eq.\nr{dissipative} for $\bfx=0$)
of the energy-momentum tensor,
$\Delta T_{ij}=-\zeta \delta_{ij}\nabla\cdot{\bf v}= -3\zeta
r'/r\delta_{ij}$ we can identify a time dependent bulk viscosity
(take $h=1$) \be \zeta(t)={N_c^2\over 24\pi^2}{r'r''\over r^2}.
\label{zeta} \ee Bulk viscosity vanishes for a conformal massless
system, but here it is precisely the anomaly which gives rise to
it. A positive $\zeta$ implies that entropy increases as can be
explicitly verified from our solution.

\section{Entropy, gravity dual}\label{entro}
The coefficient $a(t,z)$ in \nr{5dsola} can be written in the form (for $h=1$)
\be
a(t,z)={r^2\over b(t,z)}\left[\left(1-{z^2\over z_{H+}^2}\right)
\left(1-{z^2\over z_{H-}^2}\right)\right]^2,
\ee
where
\be
z_{H\pm}^2={4r^2\over rr''\pm\sqrt{4/z_0^4+(r'^2-rr'')^2}},\qquad z_{H+}\equiv z_H(t)<z_{H-}.
\label{horpos}
\ee
We have now a bulk theory with a dynamical horizon at
$z_H(t)$. The entropy density $s(t)$ then is given by the area of the horizon:
\be
s(t)r^3(t)={S\over V_3}={{\rm Area}\over 4G_5V_3}={1\over 4G_5}\int {d^3x\over V_3}
\sqrt\gamma={\CL^3b^{3/2}\over4G_5z_H^3}={N_c^2b^{3/2}(t,z_H)\over2\pi z_H^3},
\label{entropy}
\ee
where $\gamma$ is the determinant of the metric of the $(x^1,x^2,x^3)$ subspace in
\nr{bulkmetric} and
\be
{b(t,z_H)\over z_H^2}={4/z_0^4+[rr''-r'^2+\sqrt{4/z_0^4+(r'^2-rr'')^2}\,\,]^2\over
4(rr''+\sqrt{4/z_0^4+(r'^2-rr'')^2})}.
\ee

For arbitrary $r(t)$ we thus have the entropy, but the issue
of defining a temperature for a dynamical horizon is a very
complicated one \cite{ashtekar}. It should be defined so as to
satisfy the thermodynamic relations $\epsilon+p=Ts$, $s(T)=dp/dT$,
for thermal energy density and pressure.

Further, it is of interest to compute the curvature invariants for
the solution \nr{5dsol} in order to study whether the bulk
solution is singular or not $\cite{jp}$. Because
(\ref{bulkmetric}) is a solution to the equations of motion, we
automatically have ${\cal R}=-20/\CL^2$, $\CR^{MN} \CR_{MN}=80/\CL^4$. The
expression for $\CR^{MNPQ}\CR_{MNPQ}$
simplifies to the form
\be
\CR^{MNPQ}\CR_{MNPQ}={1\over\CL^4}\left\{40+72
\left[{z^2\over z_0^2b(t,z)}\right]^4\right\}, \label{r2}
\ee
where $b(t,z)$ is given by \nr{5dsol}. The 40 here is the
maximally symmetric part $2{\cal R}^2/(d^2-d)$ of
$\CR_{MNPQ}^2$.

To calibrate the above expressions note that    
for $r=1$ we have $z_H^2=2z_0^2$ and $b(z_H)/z_H^2=1/z_0^2$, which, using
\nr{entropy}, gives $s=N_c^2/(2\pi z_0^3)$, the standard value for a static horizon.
Furthermore, the metric then is \nr{bhmetric} with the Hawking temperature
$T=1/(\pi z_0)$ and
one obtains the standard result $s=\pi^2N_c^2T^3/2,
p=\pi^2N_c^2T^4/8$ for the pressure of strongly coupled thermalised supersymmetric
Yang-Mills matter. Also, for $r=1$ the curvature invariant \nr{r2} has the value
$\CR_{MNPQ}^2=112/\CL^4$ at the horizon.

Our goal is to find the gravity dual of the flow in \nr{emtensor} with a powerlike
$\epsilon(t)$ in \nr{restframeeps} with $d=4$.
For a general powerlike behavior $r= (t/t_0)^n$ we have $r'=nr/t,\,r''=n(n-1)r/t^2$ so that
\ba
\epsilon&=& {3a\over4t^4}\left({4t^4\over z_0^4r^4}+n^4\right),\qquad
T^1_1={a\over4t^4}\left({4t^4\over z_0^4r^4}-3n^4+4n^3\right)\\
z_H^2(t)&=&{4t^2\over n^2-n+\sqrt{4t^4/(z_0^4r^4)+n^2}}\\
s&=&4\pi a{1\over 8t^3}\left[{4t^4/(z_0^4r^4)+\left(-n+\sqrt{4t^4/(z_0^4r^4)+n^2}\right)^2
\over
n^2-n+\sqrt{4t^4/(z_0^4r^4)+n^2}}\right]^{3/2},
\ea
where $a\equiv N_c^2/(8\pi^2)$. All the quantities depend essentially on the
combination $4t^4/(z_0^4r^4(t))$.

The required behavior $\epsilon=\epsilon_0/t^4$ is thus obtained when $n=1$,
$r(t)=t/t_0$. Then
\ba
z_H&=&{2t\over
(4t_0^4/z_0^4+1)^{1/4}},\qquad \epsilon={3a\over 4t^4}
\left({4t_0^4\over z_0^4}+1\right)=
\frac{3\pi^2 N_c^2}{8}\bigg(\frac{\sqrt{\,2}}{\pi z_H(t)}\bigg)^4=3p,\\
s &=& {\sqrt{\,2}\,\pi a\over t^3}\left(\sqrt{{4t_0^4\over z_0^4}+1}-1\right)^{3/2}=
\frac{\pi^2 N_c^2}{2}\left(\left(\frac{\sqrt{\,2}}{\pi z_H(t)}\right
)^2-\frac{1}{2\pi^2 t^2}\right)^{3/2}.
\label{nis1}
\ea

The gravity dual we searched for thus is given by the metric \nr{5dsola},\nr{5dsol}
with $r(t)=t/t_0$. This result has the following properties:

\begin{itemize}
\item
For the boundary flow \nr{restframeeps} the scale factor $\epsilon_0$
is given by initial conditions. Here this is related to the parameter
$t_0$ of the dual metric.
Writing $1/z_0=\pi T_H$ it appears in the combination $4\pi^4 t_0^4T_H^4+1$,
where the factor $+1$ represents the effect of the curvature of the
boundary metric. We want this to be negligible, since the little bang
takes place in a flat space. Thus we have to demand
\be
\pi T_Ht_0={t_0\over z_0}\,\gsim\, 1.
\label{tcond}
\ee
Then also the scale factor in \nr{restframeeps} has to satisfy
$\epsilon_0\,\gsim\, 3a=3N_c^2/(8\pi^2)$.

\item
Since we are looking for the gravity dual of a thermalised system, we
also have to define the temperature.
This is defined by the two conditions $T=(\epsilon+p)/s=4p/s$ and $s=dp/dT$.
These can be integrated to give $p=cT^4$, $s=4cT^3$,
where $c=(s/4)^4/p^3$ is a constant, given by Eqs. \nr{nis1}.
For the temperature one obtains from $T=4p/s$ and Eqs. \nr{nis1} that
\be
\sqrt2 \pi Tt={4\pi^4t_0^4T_H^4+1\over
\left(\sqrt{4\pi^4t_0^4T_H^4+1}-1\right)^{3/2}}\rightarrow
\sqrt2 \pi T_Ht_0\quad {\rm if}\,\,\pi T_Ht_0\gsim 1.
\label{temp}
\ee
Inserting this to $s=4cT^3$, $p=cT^4$, one sees that the equation of state
$s(t)=\fr12\pi^2 N_c^2 T^3(t)$ and $p(t)=\fr18\pi^2 N_c^2 T^4(t)$,
appropriate for the matter we are considering, is obtained if
$\pi T_Ht_0=t_0/z_0\,\gsim\, 1$.
Thus matter with correct equation of state at a temperature satisfying $Tt=T_Ht_0$
is defined in the boundary theory when \nr{tcond} holds.
The behaviour $Tt\sim$ constant is
the analogue of the $Tt^{1/3}$= constant behaviour for a longitudinally expanding system.

\item
Eq. \nr{tcond} is also a natural condition for the initial temperature of
a matter system thermalised at $t_0$ at the initial temperature $T_H$,
just from the uncertainty principle.
For $t_0\,\lsim\, z_0=1/\pi T_H$ one is in a pre-thermalisation region in which
$T$, $\epsilon=3p$ and $s$ are formally defined by the above equations
but are not related by the correct equation of state.
In this pre-thermalisation region curvature effects are large.

\item
The adiabatic power $\epsilon\sim 1/t^{4/3}$ for the flow in \nr{1+1d}
was singled out in \cite{jp}
by demanding regularity of the invariant $\CR_{MNPQ}^2$ at the horizon.
The same argument works out here, too. Evaluating \nr{r2}
at the horizon for $r(t)\sim t^n$
shows that it is constant for $n=1$, grows without bounds $\sim
t^{8(n-1)}$ for $t\gg z_0$ when $n>1$ and grows without bounds
$\sim 1/t^{8(1-n)}$ for $t\ll z_0$ when $n<1$. Thus, if one demands that
$\CR_{MNPQ}^2$ at the horizon be regular when $t\to\infty$ or $t\to 0$,
one again finds that $n=1$, corresponding to $\epsilon(t)=\epsilon_0/t^4$,
is the only option.

\end{itemize}

\section{Conclusions}
The goal of this work was to develop the framework for studying
time dependent systems in the gauge/gravity duality picture in
\cite{jp, jp2,nakamura1,nakamura2, janik} by analysing an exact
solution of the 5d AdS gravity equation. The price we had to pay
was that more symmetry had to be assumed: we had as the goal 3d
radial expansion. It turns out that the boundary metric is the
standard cosmological FRW metric with an unknown time dependent
radial function $r(t)$. In cosmology one, of course, can determine
$r(t)$
from Einstein's equations when $T_\mn$ is given.
Since we want to have adiabatically expanding matter in the
boundary theory, we determine
$r(t)=t/t_0$ so that the correct $T_\mn$ for the expanding system
is obtained. The uncertainty in $t_0$ corresponds to arbitrariness
of the initial conditions of the radial flow, for $t_0T_0\gsim1$
the system is thermalised and curvature effects of the boundary
metric are negligible. Conversely, by demanding regularity of
$\CR_{MNPQ}^2$ at the horizon, one can derive the adiabatic exponent
in $\epsilon\sim 1/t^4$.

It would be most interesting to also obtain information on the transport
coefficients. However, with spherical symmetry, ${\bf v}={\bf r}/t$, shear viscosity does
not contribute, Eq. \nr{dissipative},
and for the perfect fluid case $r(t)\sim t$ the anomaly,
interpreted as bulk viscosity, vanishes.

There are clearly many issues requiring further study. For the first, an exact
solution with the symmetries \nr{jpmetric} would be very useful. The boundary
metric would then be of the form diag$(-1,g^2(t),r^2(t),r^2(t))$ \cite{nakamura2}.
For the second, consequences of the time dependent horizon and its effects on the
determination of entropy and temperature should be understood better. The
passage from known energy density and pressure, with non-thermal components,
and entropy to temperature is the key issue here. For the third,
exact solutions with scalar and form fields could possibly give more information.

\vspace{0.5cm}
Acknowledgements. We thank Esko Keski-Vakkuri for discussions and K. Skenderis,
J. Louko and R. Janik
for advice. This research has been supported by Academy of
Finland, contract number 109720.

\end{document}